\documentclass[aps, a4paper,superscriptaddress,12pt,preprintnumbers,floatfix,nofootinbib,amsmath,amssymb]{revtex4}
\usepackage{amstext,amssymb}
\usepackage{amsmath}
\usepackage{graphicx}
\usepackage{tikz}
\usepackage{tikz-feynman}
\usepackage{tikz-feynman,contour}
\usepackage[hyperfootnotes=false]{hyperref}
\usepackage{xspace}
\usepackage{color}
\usepackage{units}
\usepackage{subfigure}
\usepackage{slashed} 

\newcommand{\bea}{\begin{eqnarray}}
\newcommand{\eea}{\end{eqnarray}}
\newcommand{\nn}{\nonumber}

\global\long\def\d{\partial}

\def\s1{\hat s}

\def\U1mt{U(1)_{L_\mu-L_\tau}}

\def\A{\mathcal{A}}

\def\SO10{\text{SO}(10)}

\def \epsilon {\varepsilon} 

\usepackage{longtable}
\usepackage{needspace}
\def\SO10{\text{SO}(10)}

\newcommand{\red}[1]{\textcolor{red}{#1}}

\begin{document}
\title{\hspace*{-1.0cm} Correlating neutrino magnetic moment and scalar triplet dark matter to enlighten XENONnT bounds in a Type-II model}
\author{Shivaramakrishna \surname{Singirala}}
\email{krishnas542@gmail.com}
\affiliation{School of Physics, University of Hyderabad, Hyderabad 500046, India}
\author{Dinesh Kumar \surname{Singha}}
\email{dinesh.sin.187@gmail.com}
\affiliation{School of Physics, University of Hyderabad, Hyderabad 500046, India}
\author{Rukmani \surname{Mohanta}}
\email{rmsp@uohyd.ac.in}
\affiliation{School of Physics, University of Hyderabad, Hyderabad 500046, India}

\begin{abstract}
\vspace*{1cm}

\end{abstract}
\maketitle
We investigate neutrino magnetic moment, triplet scalar dark matter in a Type-II radiative seesaw scenario. With three vector-like fermion doublets and two scalar triplets, we provide a loop level setup for the electromagnetic vertex of neutrinos. All the scalar multiplet components constitute the total dark matter abundance of the Universe and also their scattering cross section with detector lie below the  experimental upper limit. Using the consistent parameter space in dark matter domain, we obtain light neutrino mass in sub-eV scale and also magnetic moment in the desired range. We further derive the constraints on neutrino transition magnetic moments, consistent with XENONnT limit.

\newpage
\section{Introduction}
Standard Model (SM) of particle physics has been an enduring theory, so victorious in explaining the nature at fundamental level. Despite its remarkable success  in meeting the experimental observations, it fails to explain several anomalous phenomena such as matter dominance over anti-matter, oscillation of neutrinos  and its correlation with non-zero neutrino  masses, nature and identity of dark matter, etc. Numerous extensions to the SM have been proposed to resolve these flaws, sparking a constant struggle between theorists and instrumentalists to understand the true nature of our Universe. Neutrino oscillations have been confirmed by a variety of experiments, demonstrating two unique mass squared differences coming from the solar and atmospheric sectors. Theoretical community is continually trying to understand the unique properties of neutrinos, particularly their extremely small masses. As a consequence of non-zero masses of neutrinos, many new avenues beyond the standard model (BSM) are expected to exist,  one among them is, neutrinos having electromagnetic properties such as electric and magnetic moments.
As the interaction cross section of neutrinos with matter are  extremely small, it is hard to detect them and even harder to directly measure their electromagnetic properties with the current experiments. The most practical course of action in the present situation is to set limits on  these new properties based on the available experimental data and on that note, here we focus on neutrino magnetic moment ($\nu$MM) and more specifically transition magnetic moment.

 As we detect the neutrinos indirectly, one of the best ways is to study their properties by investigating neutrino-electron elastic scattering in the detector. It is more effective to probe neutrino magnetic moment in the lower values of electron recoil energy and the experiments with low threshold and good energy resolution are suitable in this context. Solar experiments such as BOREXINO \cite{Borexino:2017fbd}, Super-Kamiokande \cite{Super-Kamiokande:2004wqk} and reactor experiments like GEMMA \cite{Beda:2012zz}, TEXONO \cite{TEXONO:2006xds}, MUNU \cite{MUNU:2003peb} are providing some competing bounds on neutrino magnetic moments. However, more stringent constraint comes from the astrophysical sources \cite{Kopp:2022cug, Jana:2023ufy, Alok:2022pdn, Ayala:2014pea, Viaux:2013lha, MillerBertolami:2014rka, Corsico:2014mpa} and the next best limit comes from the recent   XENONnT experiment \cite{XENON:2022ltv}. Other recent works on neutrino electromagnetic properties can be found in the literature \cite{A:2022acy,Khan:2022bel,Chatterjee:2022gbo, AtzoriCorona:2022jeb, deGouvea:2022znk, Dutta:2022fdt, Xu:2022wcq, Bansal:2022zpi, Li:2022dkc, Carenza:2022ngg, DeRomeri:2022twg, MammenAbraham:2023psg}. Here, we are interested in deriving an upper bound on transition magnetic moment in a minimalistic model.

Moving on, the physics of dark matter (DM)  has been the hot cake in physics community, striving hard to reveal its characteristics. So far, we only have the  estimation for its abundance from the cosmic microwave background and Planck satellite \cite{Aghanim:2018eyx} suggests its density using the parameter $\Omega h^2 \sim 0.12$. Bullet cluster system \cite{Clowe:2006eq} predicts the dark matter to be weakly interacting and eventually a WIMP (weakly interacting massive particle) with the cross section $\sigma v \sim 10^{-9}$ ${\rm GeV}^{-2}$ seems to be one of the possible strategies to match current abundance of Universe in the particle physics perspective \cite{Murayama:2007ek}. Since the interaction strength of dark matter with the visible sector is extremely small, its detection has been  like Everest climb challenge, over the decades. Only an upper limit is levied on the DM-detector cross section and several collaborations are working hard to make the bound more sensitive and stringent. So far, no direct signal of dark matter is reported despite the assiduous attempts of the experimentalists and it is always interesting to look for indirect signs. As the neutrino sector is experimentally well established and produced several compelling results in verifying neutrino oscillation parameters with high accuracy, it will be a decent choice to correlate DM with light neutrino properties and create suitable avenues of an indirect probe. 

The primary motive of this work is to provide a simple and minimal model to obtain neutrino magnetic moment in the light of dark matter. In other words, we realize the neutrino electromagnetic vertex at one-loop level, with dark matter particles running in the loop and then study neutrino and dark matter properties in a collective manner. We enrich SM with two scalar triplets (one with zero hypercharge and other with $Y=1$) and vector-like lepton doublets to design a Type-II radiative scenario. In detail, we discuss inert triplet dark matter, neutrino mass and neutrino magnetic moment in the spotlight of XENONnT. In our recent paper, we made a similar study in the context of Type-III radiative seesaw scenario \cite{Singirala:2023zos} where, scalar dark matter (admixture of two $Y=1/2$ doublets), neutrino oscillation phenomenology is investigated and a specific range of neutrino magnetic moment is achieved to explain XENON1T excess. Comparing both the works, the unique hypercharge of scalar multiplets create different gauge coupling strength (especially with $Z$ boson) and on top, scalar mixing differs. As we shall discuss below that this distinction will alter the impact of gauge mediated annihilation channels, thereby generating a unique allowed parameter space.

The paper is organized as follows. In section-II, we describe the model framework with particle content and relevant interaction terms. In section-III, we derive mass spectrum and section-IV deals with neutrino properties, while section-V narrates dark matter observables. In section-VI, we provide a detailed analysis and consistent common parameter space and also comment on oblique parameters and collider constraints (if any). Finally, the bounds on $\nu$MM using XENONnT data is discussed in section-VII.

\section{Details of Type-II radiative seesaw framework}
The primary aim of the present model is to realize neutrino electromagnetic vertex at one-loop with dark matter. In this work, we look at Type-II case by extending SM framework with three vector-like fermion doublets ($\psi_{k}$), where $k=1,2,3$ and two inert scalar triplets, one complex ($\Delta$) and the other being real ($T$). The particle content along with their charges are displayed in Table. \ref{typeii_model}. 
\begin{table}[htb]
\caption{Fields and their charges in the present model.}
\begin{center}
\begin{tabular}{|c|c|c|c|c|}
	\hline
			& Field	& $ SU(3)_C \times SU(2)_L\times U(1)_Y$	& $Z_2$\\
	\hline
	\hline
	Fermions	& $Q_L \equiv(u, d)^T_L$			& $(\textbf{3},\textbf{2},~ 1/6)$	 & $+$\\
			& $u_R$							& $(\textbf{3},\textbf{1},~ 2/3)$	&  $+$	\\
			& $d_R$							& $(\textbf{3},\textbf{1},~-1/3)$	 & $+$\\
			& $\ell_L \equiv(\nu,~e)^T_L$	& $(\textbf{1},\textbf{2},~  -1/2)$	 & $+$\\
			& $e_R$							& $(\textbf{1},\textbf{1},~  -1)$		 & $+$\\
					& $\psi_{k(L,R)}$							& $(\textbf{1},\textbf{2},~ -1/2)$		& $-$\\
\hline	
	Scalars	& $H$							& $(\textbf{1},\textbf{2},~ 1/2)$	&    $+$\\ 
	&  $\Delta$ & $(\textbf{1},\textbf{3},1)$ &   $-$ \\
&  $T$ & $(\textbf{1},\textbf{3},0)$ &   $-$ \\
			\hline
	\hline
\end{tabular}
\label{typeii_model}
\end{center}
\end{table}

The relevant Lagrangian terms of the model are given by \cite{Lu:2016dbc,Chen:2020ark,Sahoo:2021vug}
\begin{eqnarray}
\mathcal{L}_{\psi} &&= \left(y_{\alpha k} \overline{\ell^c_{\alpha L}}i\sigma_2 \Delta \psi_{kL} + y^\prime_{\alpha k} \overline{\ell_{\alpha L}} T \psi_{kR} + {\rm h.c}\right) + M_\psi \overline{\psi_{k}} \psi_{k}+{\psi_{k}}\gamma^\mu D_\mu \psi_{k}\;,
\end{eqnarray}
where, the new $SU(2)_L$ doublet in component form is $\psi_k = \begin{pmatrix}
		 \psi^0_k	\\
		 \psi^-_k	\\
	\end{pmatrix}$ and its covariant derivative is given by
\begin{eqnarray}
 D_\mu \psi_{k} = \left(\d_\mu  + \frac{i}{2} g~  \sigma_a W^a_\mu  - \frac{i}{2}g^\prime B_\mu \right)\psi_k\;,
\end{eqnarray}
where, $\sigma_a$ with $a=1,2,3$ stand for the Pauli matrices. The scalar Lagrangian takes the form 
 \begin{align}
\mathcal{L}_{\rm scalar} &= (D_\mu \Delta)^\dagger (D^\mu \Delta) +\frac{1}{2}(D_\mu T)^\dagger (D^\mu T) - V\;,
\end{align}
where, the inert triplets are denoted by $\Delta = \begin{pmatrix}
		 \Delta^+/\sqrt{2} & \Delta^{++}		\\
		 \Delta^0	& -\Delta^+/\sqrt{2}\\
	\end{pmatrix}$, with $\Delta^0 = \displaystyle{\frac{\Delta^0_R+i\Delta^0_I}{\sqrt{2}}}$ and $T = \begin{pmatrix}
		 T^0/\sqrt{2} & T^{+}		\\
		 T^-	& -T^0/\sqrt{2}\\
	\end{pmatrix}$. In the above, the covariant derivatives are given by 
\begin{eqnarray}
&&D_\mu \Delta= \d_\mu \Delta + ig \left[\sum_{a=1}^3 \frac{\sigma^a}{2}W^a_\mu, \Delta\right] + i g^\prime B_\mu \Delta\;,\nn\\
&&D_\mu T= \d_\mu T + ig \left[\sum_{a=1}^3 \frac{\sigma^a}{2}W^a_\mu, T\right].
\end{eqnarray}
The scalar potential takes the form	
\begin{align}
V &=  \mu^2_H  H^\dagger H + \lambda_H (H^\dagger H)^2 +  \mu^2_\Delta {\rm Tr} (\Delta^\dagger \Delta) + \frac{\mu^2_T}{2} {\rm Tr} (T^\dagger T) +\lambda_\Delta {\rm Tr}(\Delta^\dagger \Delta\Delta^\dagger \Delta)  + \lambda^\prime_\Delta {\rm Tr}(\Delta^\dagger \Delta)^2 \nn\\   
      & +\frac{\lambda_T}{4} {\rm Tr} (T^\dagger T)^2 + \lambda_{H\Delta} (H^\dagger \Delta \Delta^\dagger H) + \lambda^\prime_{H\Delta} (H^\dagger H) {\rm Tr}(\Delta^\dagger \Delta)  + \frac{\lambda_{HT}}{2} (H^\dagger H) {\rm Tr}(T^\dagger T) 
       \nonumber\\ & + \frac{\lambda_{\Delta T}}{2} {\rm Tr}(\Delta^\dagger \Delta) {\rm Tr}(T^\dagger T)+ \frac{\lambda_{H\Delta T}}{2} (H^T \tilde{\Delta}T H + {\rm h.c}).     
\label{scalarpot}
\end{align} 
\section{Mass mixing in scalar sector}
The mass matrices of the charged and neural components are given by
\begin{align}
{\mathcal M}^2_{C} =
	\begin{pmatrix}
		 \Lambda_{T^+}	& -\lambda_{H\Delta T}\frac{v^2}{4\sqrt{2}}	\\
		 -\lambda_{H\Delta T}\frac{v^2}{4\sqrt{2}}	& \Lambda_{\Delta^+}	\\
	\end{pmatrix}, \quad
{\mathcal M}^2_R =
	\begin{pmatrix}
		 \Lambda_{T^0}	& \lambda_{H\Delta T}\frac{v^2}{4}\\
		 \lambda_{H\Delta T}\frac{v^2}{4}	& \Lambda_{\Delta^0_R}\\
	\end{pmatrix}.
 \label{scalar_matrix}
	\end{align}
Here, 
\begin{eqnarray}
&&\Lambda_{T^+} = \mu_{T}^2  +  \lambda_{HT}\frac{v^2}{2}\;,\nn\\
&&\Lambda_{\Delta^+} = \mu_{\Delta}^2  + (\lambda_{H\Delta}+2\lambda^\prime_{H\Delta})\frac{v^2}{4}\;,\nn\\
&&\Lambda_{T^0} = \mu_{T}^2  + \lambda_{HT}\frac{v^2}{2}\;,\nn\\
&&\Lambda_{\Delta^0_R} = \mu_{\Delta}^2  + (\lambda_{H\Delta}+\lambda^\prime_{H\Delta})\frac{v^2}{2}.
\end{eqnarray}
One can diagonalize the above mass matrices using
$
U_{\theta_{C,R}}
	=
	\begin{pmatrix}
		 \cos{\theta_{C,R}}	& \sin{\theta_{C,R}}	\\
		 -\sin{\theta_{C,R}}	& \cos{\theta_{C,R}}	\\
	\end{pmatrix}
$ as
\begin{eqnarray}
&& U_{\theta_C}^T {\mathcal M}^2_{C} U_{\theta_C} = {\rm{diag}}(M^2_{{C1}},M^2_{{C2}}) \quad {\rm with} \quad
\theta_C = \tan^{-1}\left[\frac{-\lambda_{H\Delta T}v^2}{2\sqrt{2}(\Lambda_{\Delta^+}-\Lambda_{T^+})}\right],\nn\\
&&U_{\theta_R}^T {\mathcal M}^2_{R} U_{\theta_R} = {\rm{diag}}(M^2_{{R1}},M^2_{{R2}}) \quad {\rm with} \quad
\theta_R = \tan^{-1}\left[\frac{\lambda_{H\Delta T}v^2}{2(\Lambda_{\Delta^0_R}-\Lambda_{T^0})}\right].
\end{eqnarray}
The flavor and mass eigenstates can be related as
\begin{align}
	\begin{pmatrix}
		 T^+\\
		 \Delta^+\\
	\end{pmatrix} =
	U_{\theta_C}
		\begin{pmatrix}
		 \phi^+_1\\
		 \phi^+_2\\
	\end{pmatrix}, \quad
	\begin{pmatrix}
		 T^0\\
		 \Delta^0_R\\
	\end{pmatrix} =
	U_{\theta_R}		\begin{pmatrix}
		 \phi^0_{1R}\\
		 \phi^0_{2R}\\
	\end{pmatrix},
 \label{flav to mass}
\end{align}
The masses of doubly charged and CP-odd scalar follow as
\begin{eqnarray}
&&M^2_{CC} = \mu_{\Delta}^2  + \lambda^\prime_{H\Delta}\frac{v^2}{2},\nn\\
&&M^2_{I} = \mu_{\Delta}^2  + (\lambda_{H\Delta}+\lambda^\prime_{H\Delta})\frac{v^2}{2}.
\end{eqnarray} 
In fermion sector, one-loop electroweak radiative corrections provide a mass splitting of $166$ MeV \cite{Cirelli:2005uq,Ma:2008cu} between neutral and charged components of $\psi$. We work in the high scale regime of $\psi$ and so we take $M_{\psi^+} \simeq M_{\psi^0} = M_{\psi}$.
\section{Neutrino phenomenology}
\subsection{Neutrino Magnetic moment}
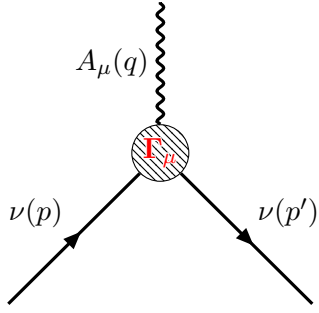
\begin{figure}
  \centering
  \begin{tikzpicture}
    \begin{feynman}
      \vertex (a) at (-2,-2);
      \vertex (b) at ( 0,2) ;
      \vertex (c) at (2, -2) ;
      \vertex[blob] (m) at ( 0, 0) {\contour{white}{\red{$\mathbf{\Gamma_{\mu}}$}}};
      \diagram* {
        (a) -- [fermion, very thick, edge label=$\nu (p)$] (m) -- [fermion, very thick, edge label=$\nu (p')$] (c),
        (b) -- [boson, very thick, edge label'=$A_{\mu} (q)$] (m),
      };
    \end{feynman}
  \end{tikzpicture}
  \caption{Effective electromagnetic vertex, where $q = p-p^\prime$.  }
  \label{emvtx_feyn}
\end{figure}

Though neutrino is electrically neutral, it can have electromagnetic interaction at loop level, as shown in Fig. \ref{emvtx_feyn}. The effective Lagrangian takes the form \cite{Xing:2011zza}
\begin{equation}
\mathcal{L}_{\rm EM} =  \overline{\nu}\Gamma_\mu \nu A^\mu.
\end{equation}
In the above, the electromagnetic vertex function can accommodate charge, electric dipole, magnetic dipole and anapole moments and varies with the type of neutrinos i.e., Dirac or Majorana. In our work, we stick to Majorana neutrino magnetic dipole moment. In general, the electromagnetic contribution to neutrino magnetic moment can be written as 
\begin{eqnarray}
    \mathcal{L} \supset \mu_{\alpha \beta} \overline{\nu_{\alpha}} \sigma^{\mu \nu} \nu_{\beta} F_{\mu \nu}  &=& \mu_{\alpha \beta} \left( \overline{\nu_{\alpha L}} \sigma^{\mu \nu} \nu^{c}_{\beta L} + \overline{\nu^{c}_{\alpha L}} \sigma^{\mu \nu} \nu_{\beta L} \right) F_{\mu \nu}\;.
\end{eqnarray}
Here, $\sigma^{\mu \nu}$ is the anti-symmetric matrix and $F_{\mu \nu}$ is the electromagnetic field strength tensor. By considering the anti-symmetric nature of fermion fields and the characteristics of the charge-conjugation matrix, one can write
\begin{eqnarray}
    \overline{\nu_{\alpha}} \sigma^{\mu \nu} \nu_{\beta} &=& - \overline{\nu_{\beta}} \sigma^{\mu \nu} \nu_{\alpha}.
\end{eqnarray}
Hence, Majorana neutrinos can have only transition (off-diagonal) magnetic moments.
In the present model, the transition magnetic moment arises from one-loop diagram shown in Fig. \ref{numag_feyn} and the expression takes the form \cite{Babu:2020ivd}
\begin{equation}
\mu_{\nu_{e\mu}} =  \sum_{k=1}^3\frac{y^\prime_{\alpha k}~ y_{\beta k}} {16\pi^{2}}M_{\psi^+_k} \cos 2\theta_C \sin 2\theta_C \left[\frac{1}{M^{2}_{C2}} \left(\log \left[\frac{M^{2}_{C2}}{M^2_{\psi^+_k}}\right]-1\right) - \frac{1}{M^{2}_{C1}} \left(\log \left[\frac{M^{2}_{C1}}{M^2_{\psi^+_k}}\right]-1\right) \right].
\end{equation}
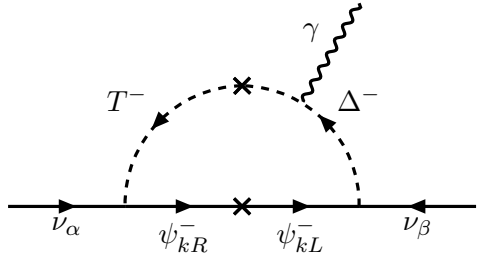
\begin{figure}
    \begin{tikzpicture}
        \begin{feynman}
            \vertex (a);
            \vertex[right=4em of a] (b);
            \vertex[right=4em of b] (c);
            \vertex[right=4em of c] (e);
            \vertex[right=4em of e] (f);
            \vertex at ($(b)!0.5!(e)!1.6cm!+90:(e)$) (d);
            \vertex at ($(b)!0.5!(e)!1.6cm!+60:(e)$) (g);
            \vertex[above=7em of e] (h);

            \diagram* {
      (a) -- [fermion, very thick, edge label'=$\nu_{\alpha}$] (b),
      (e) -- [charged scalar, very thick, quarter right, edge label'=$\Delta^-$] (d) -- [charged scalar, very thick, quarter right, insertion=0, edge label'=$T^-$] (b),
      (b) -- [fermion, very thick, edge label'=$\psi^-_{kR}$] (c),
      (c) -- [fermion, very thick, insertion=0, edge label'=$\psi^-_{kL}$] (e),
      (f) -- [fermion, very thick, edge label=$\nu_{\beta}$] (e),
      (g) -- [boson, very thick, edge label=$\gamma$] (h),
     
    };
        \end{feynman}
    \end{tikzpicture}
    \caption{One-loop Feynman diagram for transition magnetic moment.}
    \label{numag_feyn}
\end{figure}
We shall discuss bounds predicted by XENONnT on transition magnetic moment in the upcoming section.
\subsection{Neutrino mass}
\begin{figure}
    \begin{tikzpicture}
        \begin{feynman}
            \vertex (a);
            \vertex[right=4em of a] (b);
            \vertex[right=4em of b] (c);
            \vertex[right=4em of c] (e);
            \vertex[right=4em of e] (f);
            \vertex at ($(b)!0.5!(e)!1.6cm!+90:(e)$) (d);
            \vertex at ($(b)!0.5!(e)!1.6cm!+60:(e)$) (g);
            \vertex[above=7em of e] (h);

            \diagram* {
      (a) -- [fermion, very thick, edge label'=$\nu_{\alpha}$] (b),
      (e) -- [charged scalar, very thick, quarter right, edge label'=$\Delta^- (\Delta^0)$]
      (d) -- [charged scalar, very thick, quarter right, insertion=0, edge label'=$T^- (T^0)$] (b),
      (b) -- [fermion, very thick, edge label'=$\psi^{-(0)}_{kR}$] (c),
      (c) -- [fermion, very thick, insertion=0, edge label'=$\psi^{-(0)}_{kL}$] (e),
      (f) -- [fermion, very thick, edge label=$\nu_{\beta}$] (e),
     
    };
        \end{feynman}
    \end{tikzpicture}
    \caption{One-loop diagram give rise to light neutrino mass.}
    \label{numass_feyn}
\end{figure}
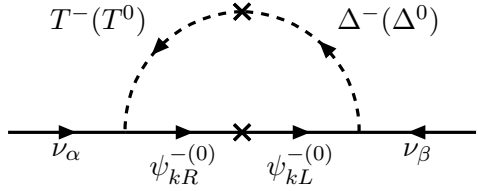

From various oscillation experiments, we know that neutrinos indeed oscillate in flavor  and posses sub-eV scale mass. In the present model, neutrino mass can arise at one-loop level, as shown in Fig. \ref{numass_feyn} with vector-like leptons and scalar triplets running in the loop. The contribution takes the form \cite{Lu:2016dbc,Chen:2020ark,Sahoo:2021vug}
\begin{eqnarray}
&&{\cal M}_{\nu_{\alpha\beta}} = \sum_{k=1}^3 \frac{y^\prime_{\alpha k} ~y_{\beta k}}{32 \pi^{2}}\sin\theta_C \cos\theta_C M_{\psi^+_k}
 \left[\frac{M_{C2}^2}{M_{\psi^+_k}^{2}-M_{C2}^2}\ln\left(\frac{M^2_{\psi^+_k}}{M^2_{C2}}\right) - \frac{M_{C1}^2}{M_{\psi^+_k}^{2} -  M_{C1}^2}\ln\left(\frac{M^2_{\psi^+_k}}{M^2_{C1}}\right)  \right] \nonumber\\
&&~~ +\sum_{k=1}^3 \frac{y^\prime_{\alpha k} ~y_{\beta k}}{32 \pi^{2}}\sin\theta_R \cos\theta_R M_{\psi^0_k}
 \left[\frac{M_{R2}^2}{M_{\psi^0_k}^{2}-M_{R2}^2}\ln\left(\frac{M^2_{\psi^0_k}}{M^2_{R2}}\right) - \frac{M_{R1}^2}{M_{\psi^0_k}^{2} -  M_{R1}^2}\ln\left(\frac{M^2_{\psi^0_k}}{M^2_{R1}}\right)  \right].
\label{nu-mass}
\end{eqnarray}

\section{Dark matter phenomenology}
\subsection{Relic density}
The neutral and charged components of scalar triplets contribute to total relic density of dark matter in the Universe. The channels include several annihilation and co-annihilation processes, mediated through SM bosons. 
The cross sections of all the viable channels is used to compute the relic density by
\begin{equation}
\label{eq:relicdensity}
\Omega h^2 = \frac{1.07 \times 10^{9} ~{\rm{GeV}}^{-1}}{ {g_\ast}^{1/2}\; M_{\rm{Pl}} }\frac{1}{J(x_f)}\;,
\end{equation}
where, the Planck mass $M_{\rm{Pl}}=1.22 \times 10^{19} ~\rm{GeV}$ and total effective relativistic degrees of freedom $g_\ast = 106.75$. The function $J$ is given by \cite{Griest:1990kh}
\begin{equation}
J(x_f)=dx\int_{x_f}^{\infty} \frac{ \langle \sigma_{\rm eff} v \rangle (x)}{x^2}\;,
\end{equation}
where, the thermally averaged cross section $\langle \sigma_{\rm eff} v \rangle$ is computed by \cite{Edsjo:1997bg} 
\begin{equation}
    \langle \sigma_{\rm eff}v \rangle = \frac{\int_0^\infty dp_{\rm eff}~ p^2_{\rm eff}W_{\rm eff} K_1 \left(\frac{\sqrt{s}}{T}\right)}{m_1^4 T\left[\sum_i \frac{g_i}{g_1}\frac{m_i^2}{m_1^2}K_2\left(\frac{m_i}{T}\right)\right]^2}\;.
\end{equation}
Here,
\begin{equation}
    W_{\rm eff} = \sum_{ij} \frac{p_{ij}}{p_{11}}\frac{g_ig_j}{g_1^2} W_{ij},
\end{equation}
with 
\begin{equation}
W_{ij} = 4p_{ij}\sqrt{s}\sigma_{ij}, \quad p_{ij} = \left(\frac{(s-(m_i+m_j)^2)(s-(m_i-m_j)^2)}{4s}\right)^\frac{1}{2}, \quad p_{\rm eff} = p_{11}.
\end{equation}
In the above, $K_1$, $K_2$ represent the modified Bessel functions. $g_i$ and $m_i$ correspond to the internal degrees of freedom and masses of particles participating annihilation/co-annihilation respectively, with $g_1$, $m_1$ pointing to the lightest one. We shall make it clear in the next section, the inert scalars of the model and their mass spectrum to discuss in detail on the specific channels that provide reasonable contribution to relic density.
\subsection{Direct detection}
The scalar dark matter can provide spin-independent (SI) scattering cross section with nucleons via Higgs boson. The effective interaction Lagrangian takes the form 
\begin{equation}
    \mathcal{L}_{\rm eff} \supset  a_q \phi^0_R\phi^0_R q\overline{q}\;, 
\end{equation}
\begin{equation}
 {\rm where} \quad a_q = \frac{1}{M_h^2 M_{R1}}M_q \left((\lambda_{H\Delta}+\lambda^\prime_{H\Delta})\sin^2\theta_R+ \frac{1}{2} \lambda_{HT}\cos^2\theta_R - \frac{1}{4} \lambda_{H\Delta T} \cos\theta_R \sin\theta_R\right).\nonumber
\end{equation}
The resulting DM-nucleon cross section is given by
\begin{equation}
    \sigma_{\rm SI} = \frac{1}{4\pi} \mu_r^2 \left(\frac{(\lambda_{H\Delta}+\lambda^\prime_{H\Delta})\sin^2\theta_R+ \frac{1}{2} \lambda_{HT}\cos^2\theta_R - \frac{1}{4} \lambda_{H\Delta T} \cos\theta_R \sin\theta_R}{M_h^2 M_{R1}}\right)^2 f_n^2 M_n^2\;,
\end{equation}
where, $f_n \sim 0.3$ \cite{Ellis:2000ds} is the Higgs-nucleon matrix element, $\mu_r = \left(\frac{M_n  M_{R1}}{M_n + M_{R1}}\right)$ is the reduced mass with $M_n$ being the nucleon mass.  In $Z$-portal, large  WIMP-nucleon cross section can arise due to $Y=1$ triplet component in dark matter admixture. However, $Z$-exchange kinematics can be forbidden by choosing the mass splitting between CP-even and CP-odd components above $\sim 100$ KeV \cite{Hambye:2009pw,LopezHonorez:2006gr}. We shall address this point in the next section with an illustrative plot.

We have used the packages LanHEP \cite{Semenov:1996es} and micrOMEGAs \cite{Pukhov:1999gg, Belanger:2006is, Belanger:2008sj} to extract dark matter relic density and also SI DM-nucleon cross section. A detailed discussion of dark matter observables with suitable plots will be discussed in the next section. 
\section{Numerical Analysis}
Here we illustrate the analysis of both neutrino and dark matter aspects in a correlative manner. There are two CP-even and two singly charged scalars that mix, in order to make the analysis simpler, we consider the mass of the one CP-even scalar ($M_{R1}$) and two mass splittings ($\delta$ and $\delta_{\rm CR}$) to derive the masses of the other CP-even and singly charged scalars. The relations are as follows 
\begin{eqnarray}
&& M_{R2} - M_{R1} = M_{C1}-M_{C2} = \delta\;, \nn\\
&& M_{C2} = M_{R1} + \delta_{\rm CR}, M_{C1} = M_{R2} + \delta_{\rm CR}\;.
\label{splitting relations}
\end{eqnarray}
One can notice the difference in mass ordering, $M_{R1} < M_{R2}$, while $M_{C1} > M_{C2}$, which comes due to the relative opposite sign in the mass matrices of Eqn. \ref{scalar_matrix}. We run the scan over model parameters in range given below 
\begin{eqnarray}
&&10 ~{\rm GeV} \le M_{R1} \le 2000 ~{\rm GeV}, \quad 0 \le \sin\theta_R, \sin\theta_C \le 1, \nn\\
&&0.1 ~{\rm GeV} \le \delta < 200 ~{\rm GeV}, \quad 0.1 ~{\rm GeV} \le \delta_{\rm CR} \le 20 ~{\rm GeV}.
\end{eqnarray}
\begin{figure}[htb!]
\centering
\includegraphics[scale=0.45]{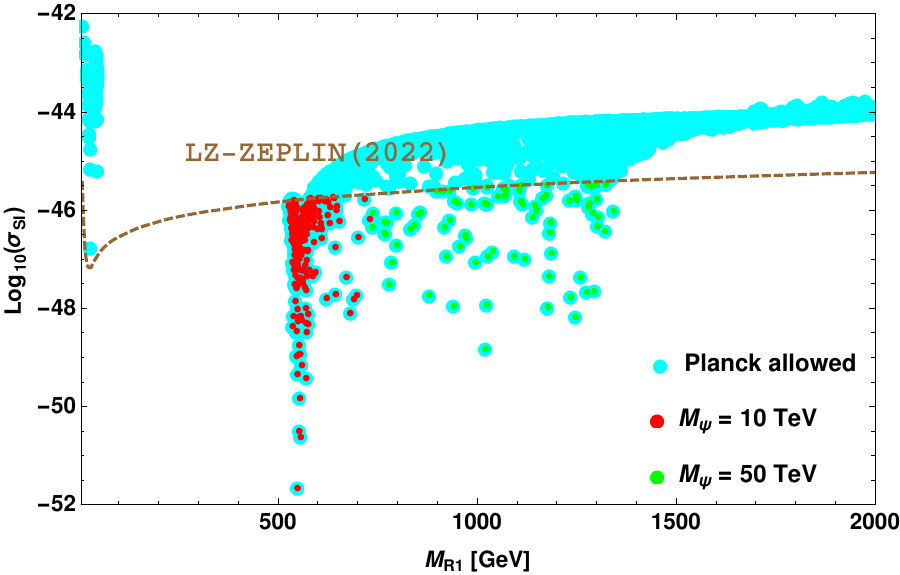}
\hspace{0.2 cm}
\includegraphics[scale=0.5]{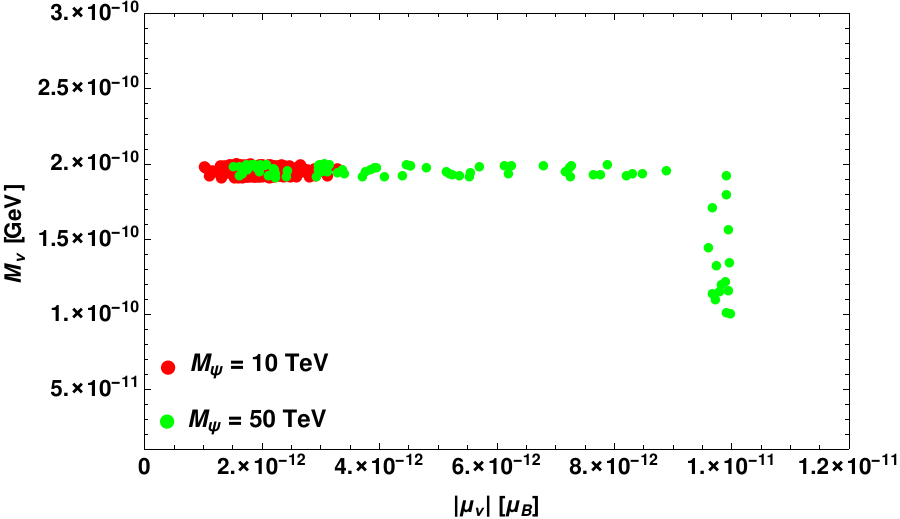}
 \caption{Left panel depicts the SI DM-nucleon cross section for the data (cyan) that satisfies Planck limit on relic density. Horizontal dashed line corresponds to LUX-ZEPLIN bound \cite{LZ:2022ufs}. Various colored data points of red and green satisfy neutrino magnetic moment and mass for specific set of values for $M_\psi$, shown  in the right panel.}
 \label{DDSI}
\end{figure}
\begin{figure}[htb!]
\centering
\includegraphics[scale=0.52]{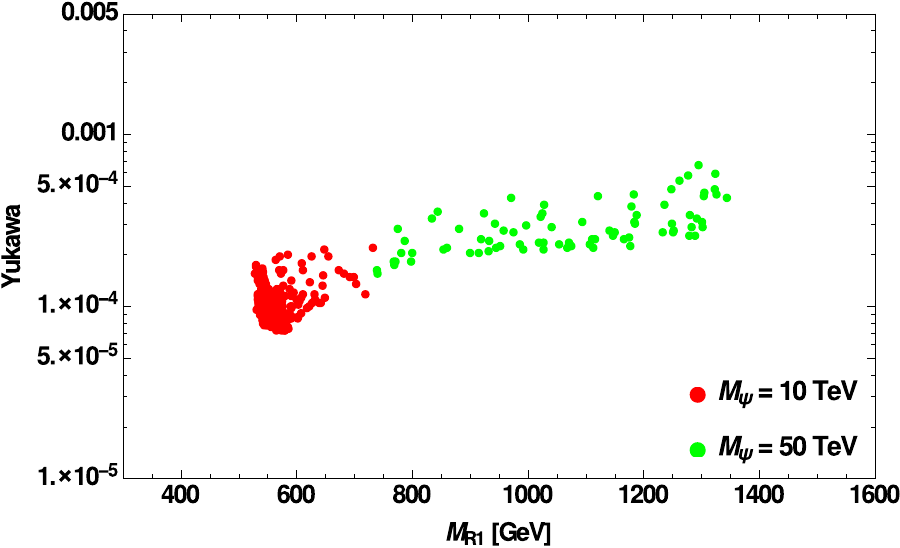}
\hspace{0.2 cm}
\includegraphics[scale=0.42]{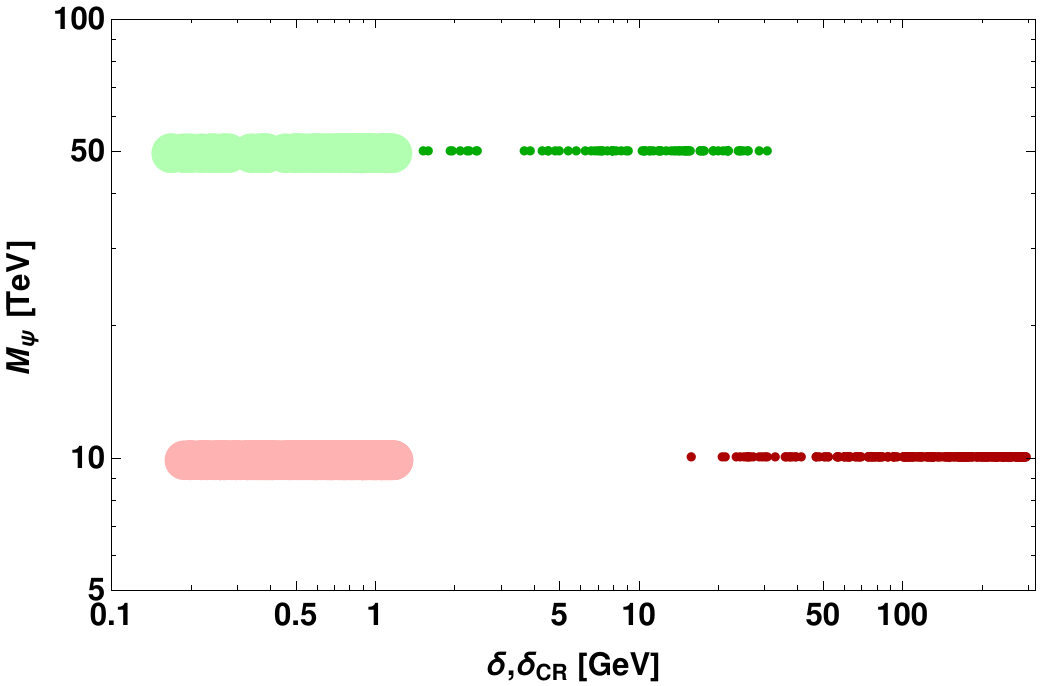}
\caption{Allowed region of Yukawa with dark matter mass for two set of values for $M_{\psi}$ in the left panel. Right panel provides the favourable region for mass splittings, where the thick bands correspond to $\delta_{\rm CR}$ and thin bands indicate $\delta$.} 
\label{neu_pheno}
\end{figure}

We first filter out the parameter space using the $3\sigma$ constraint on relic density. We then compute the DM-nucleon cross section and project it as a function of $M_{R1}$ in the left panel of Fig. \ref{DDSI}. Here, all the cyan colored data points satisfy Planck satellite data \cite{Aghanim:2018eyx} in $3\sigma$ region and the parameter space (red and green) below LZ-ZEPLIN bound \cite{LZ:2022ufs} (orange dashed line) can satisfy neutrino magnetic moment and mass in the desired range when suitable values are assigned to Yukawa and vector-like lepton mass. The obtained neutrino observables are projected in the right panel for various set of values for $M_{\psi}$. One can notice from the left panel that specific regions of DM mass are allowed for different values of heavy fermion mass. This restriction comes from neutrino magnetic moment and mass, where a particular region for DM mass and Yukawa is favoured based on the value of $M_{\psi}$ in order to satisfy the experimental bounds in neutrino sector. The same is made transparent in the left panel of Fig. \ref{neu_pheno}, right panel depicts the allowed range for mass splittings (thick bands correspond to $\delta_{\rm CR}$ and thin bands indicate $\delta$). 

\begin{table}[htb]
\caption{Set of benchmarks from the consistent parameter space.}
\begin{center}
\begin{tabular}{|c|c|c|c|c|c|c|c|c|c||c|c|}
	\hline
			& $M_{R1}$ [GeV]	& $\delta$ [GeV] & $\delta_{\rm CR}$ [GeV] & $M_{\psi}$ [TeV]& Yukawa & $\sin\theta_R$ & $\sin\theta_C$\\
	\hline
	benchmark - 1 & $580$	& $276$	& $1.17$ & $10$ & $10^{-4.15}$ & $0.083$ & $0.998$\\
	\hline
	benchmark - 2 & $1326$	& $2.28$	& $1.17$ & $50$ & $10^{-3.36}$ & $0.803$ & $0.932$\\
	\hline
\end{tabular}
\label{tab_benchmark}
\end{center}
\end{table}
\begin{table}[htb]
\caption{Neutrino and dark matter observables for the given benchmarks.}
\begin{center}
\begin{tabular}{|c|c|c|c|c|c|c|c|c|c||c|c|}
	\hline
			& $|\mu_\nu| \times 10^{12}$ [$\mu_B$]& $\mathcal{M}_{\nu}\times 10^{10}$ [GeV]& ${\rm Log}^{[\sigma_{\rm SI}]}_{10}$ ${\rm cm}^{-2}$ & $\Omega {\rm h}^2$\\
	\hline
	benchmark - 1 & $1.293$ & $1.96$ & $-47.79$ & $0.119$\\
	\hline
	benchmark - 2 & $1.624$ & $1.92$ & $-45.48$ & $0.12$\\
	\hline
\end{tabular}
\label{tab_observables}
\end{center}
\end{table}

\begin{figure}[htb!]
\centering
\includegraphics[scale=0.65]{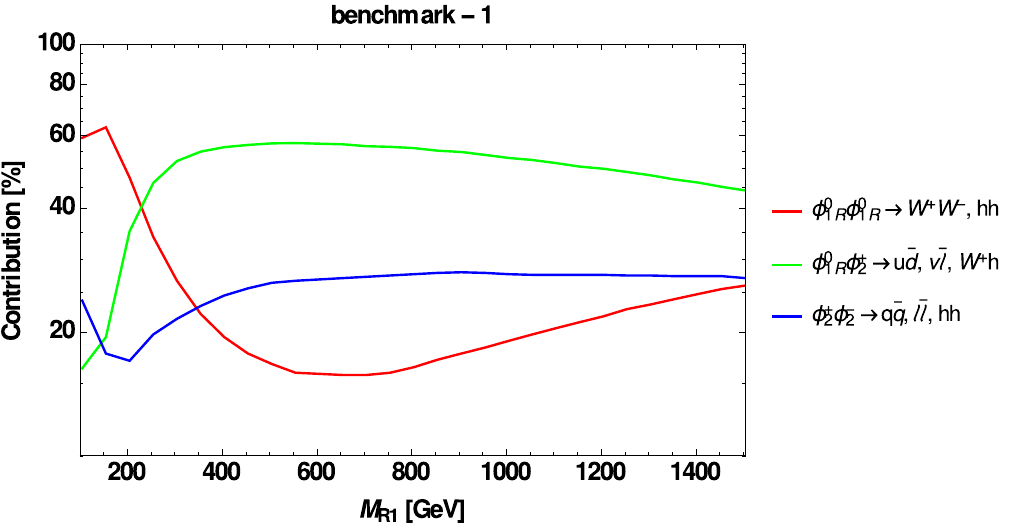}
\includegraphics[scale=0.65]{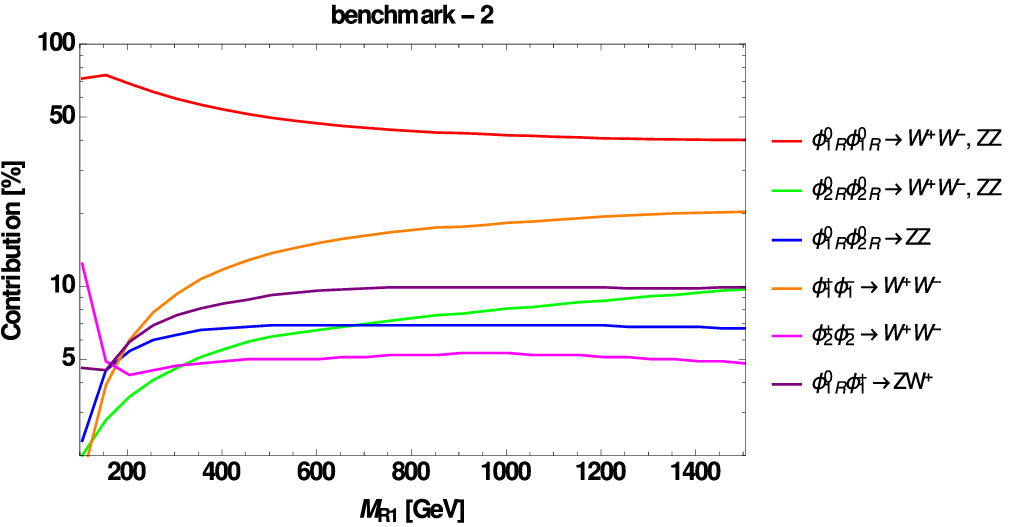}
\caption{Contribution of various annihilation and co-annihilation channels for high (upper panel) and low (lower panel) values of mass splitting $\delta$.
}
 \label{contribution plot}
\end{figure}

\begin{figure}[thb]
\begin{center}
\includegraphics[width=0.3\linewidth]{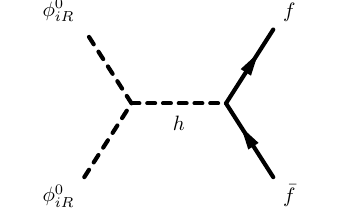}
\includegraphics[width=0.3\linewidth]{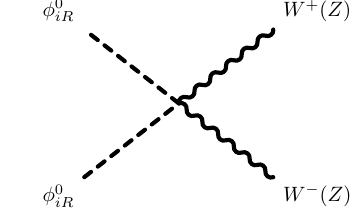}
\includegraphics[width=0.3\linewidth]{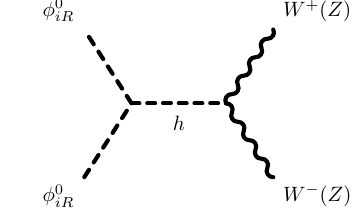}
\includegraphics[width=0.3\linewidth]{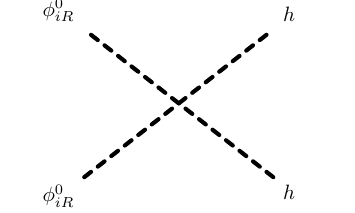}
\includegraphics[width=0.3\linewidth]{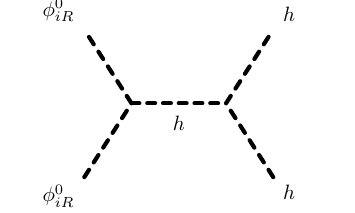}

\includegraphics[width=0.3\linewidth]{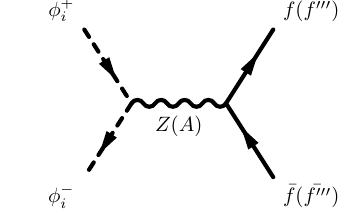}
\includegraphics[width=0.3\linewidth]{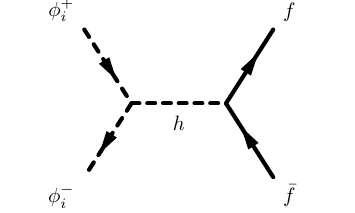}
\includegraphics[width=0.3\linewidth]{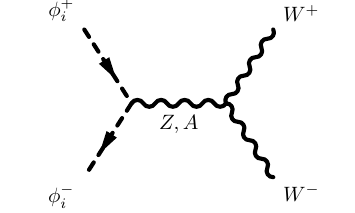}
\includegraphics[width=0.3\linewidth]{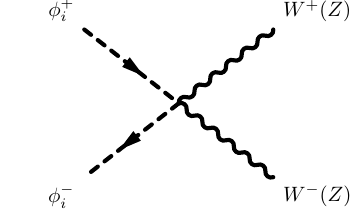}
\includegraphics[width=0.3\linewidth]{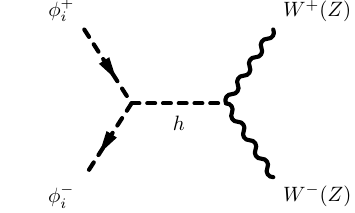}
\caption{Relevant annihilation channels contributing to relic density (classified based on output particles), where $f$ stands for SM fermions, $f^{\prime\prime\prime}$ represents SM quarks and leptons except neutrinos and $i=1,2$.}
\label{ann_feyn}
\end{center}
\end{figure}

\begin{figure}[thb]
\begin{center}
\includegraphics[width=0.3\linewidth]{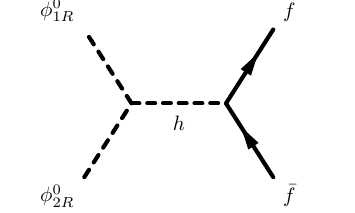}
\includegraphics[width=0.3\linewidth]{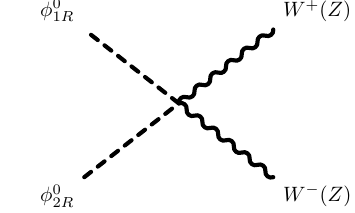}
\includegraphics[width=0.3\linewidth]{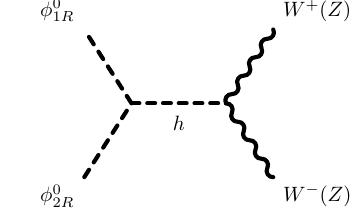}
\includegraphics[width=0.3\linewidth]{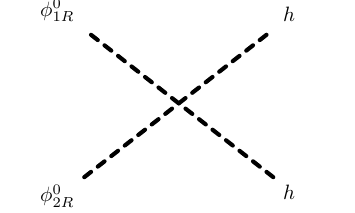}
\includegraphics[width=0.3\linewidth]{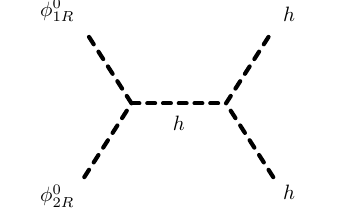}

\includegraphics[width=0.3\linewidth]{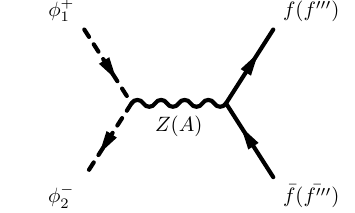}
\includegraphics[width=0.3\linewidth]{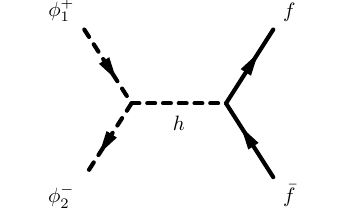}
\includegraphics[width=0.3\linewidth]{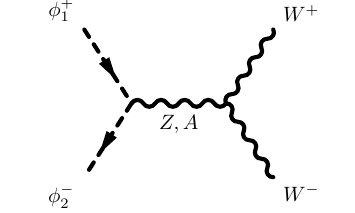}
\includegraphics[width=0.3\linewidth]{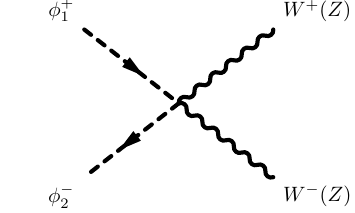}
\includegraphics[width=0.3\linewidth]{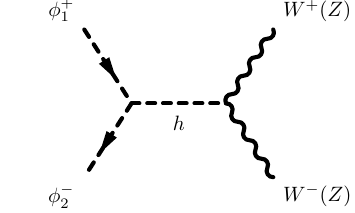}
\includegraphics[width=0.3\linewidth]{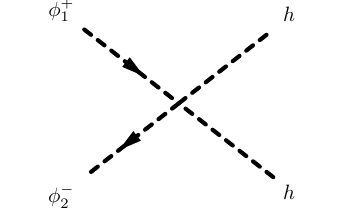}
\includegraphics[width=0.3\linewidth]{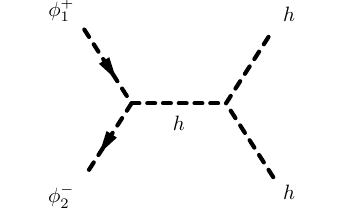}

\includegraphics[width=0.3\linewidth]{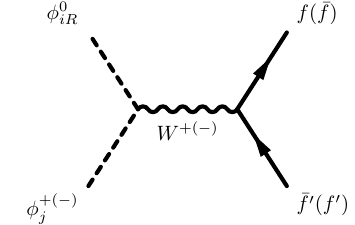}
\includegraphics[width=0.3\linewidth]{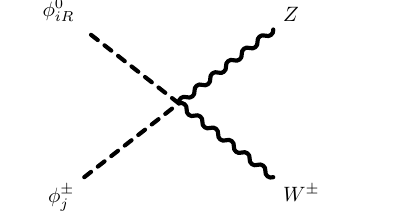}
\includegraphics[width=0.3\linewidth]{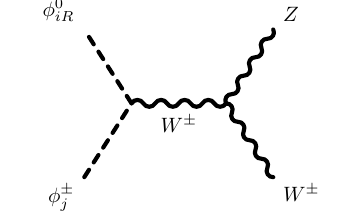}
\includegraphics[width=0.3\linewidth]{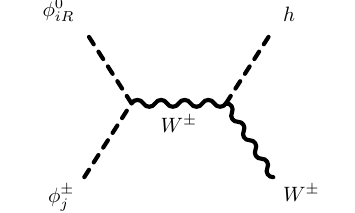}
\caption{Relevant co-annihilation channels contributing to relic density, where $f = u,c,t,e,\mu,\tau$, $f^\prime = d,s,b,\nu_e,\nu_\mu,\nu_\tau$ and $i,j = 1,2$.}
\label{coann_feyn}
\end{center}
\end{figure}

We now elaborate on the channels that provide significant contribution to relic density. The main parameters to look up are the mass splittings i.e., $\delta$ and $\delta_{\rm CR}$, who can alter the size of contribution of various channels towards DM abundance. The allowed region of these splittings can be noticed from the right panel of Fig. \ref{neu_pheno} as (in GeV scale) $0.17 < \delta_{\rm CR} < 1.22$ and $1.08 < \delta < 297$, where $\delta_{\rm CR}$ has a narrow allowed regime and small in magnitude while $\delta$ varies in a wider range. Now, we illustrate the impact of $\delta$ on relic density by choosing two extreme values in benchmarks (given in Table. \ref{tab_benchmark}, Table. \ref{tab_observables}) from the parameter space consistent with both neutrino and dark matter sectors. First with large $\delta$, the smallness of $\delta_{\rm CR}$ always makes $\phi_2^+$ nearly a mass degenerate of $\phi^0_{1R}$ and in consequence, annihilation channels $\phi^0_{1R}\phi^0_{1R} \to W^+W^-,hh$ and $\phi_2^+  \phi_2^- \to q\bar{q},\ell\bar{\ell},hh$, co-annihilations $\phi^0_{1R} \phi_2^+ \to u\bar{d},\nu\bar{\ell}, W^+h$ contribute to relic density (made apparent in the upper panel of Fig. \ref{contribution plot}). Secondly, lower value of $\delta$ creates proximity in the masses of $\phi^0_{1R}$, $\phi^0_{2R}$ and also $\phi^+_{1}$, $ \phi^+_{2}$, thereby inducing additional channels such as $\phi^0_{2R}\phi^0_{2R} \to W^+W^-,ZZ$, $\phi^0_{1R}\phi^0_{2R} \to ZZ$,  $\phi^+_{1}\phi^-_{1} \to W^+W^-$ contribute to relic density (visible in the lower panel of Fig. \ref{contribution plot}). Feynman diagrams for all these channels is displayed in Fig. \ref{ann_feyn} and Fig. \ref{coann_feyn}.

\begin{figure}[htb!]
\centering
\includegraphics[scale=0.62]{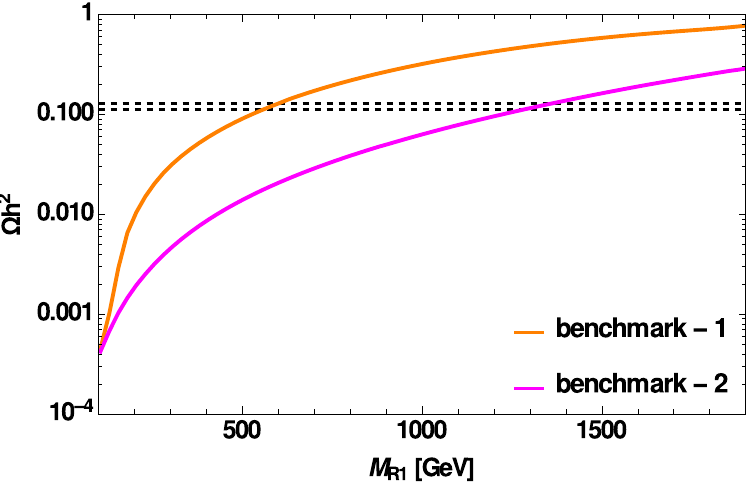}
\includegraphics[scale=0.45]{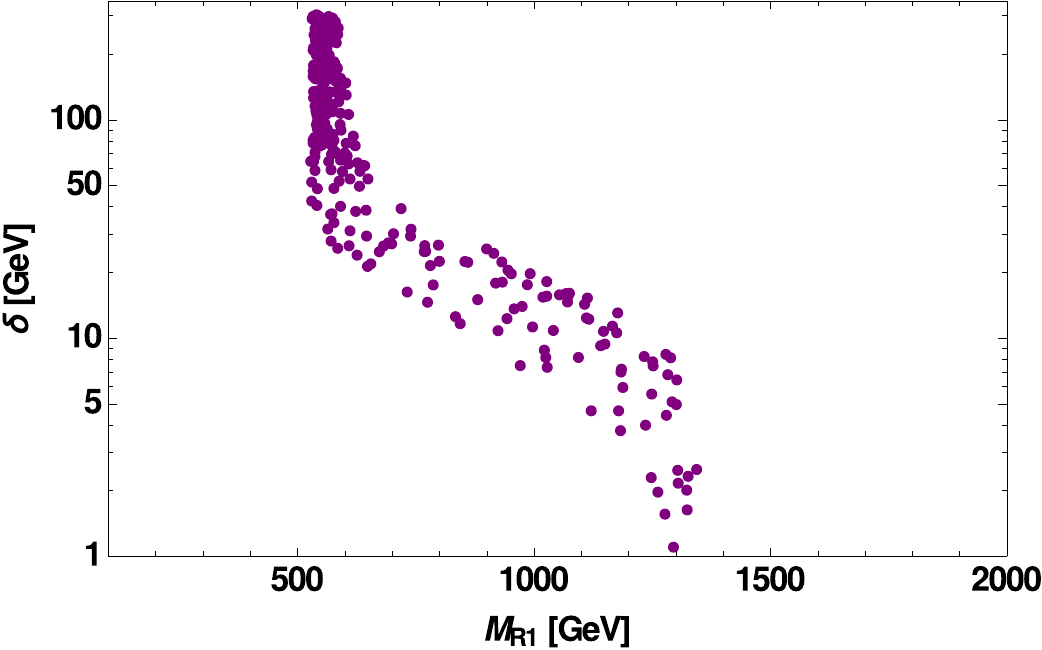}
\includegraphics[scale=0.62]{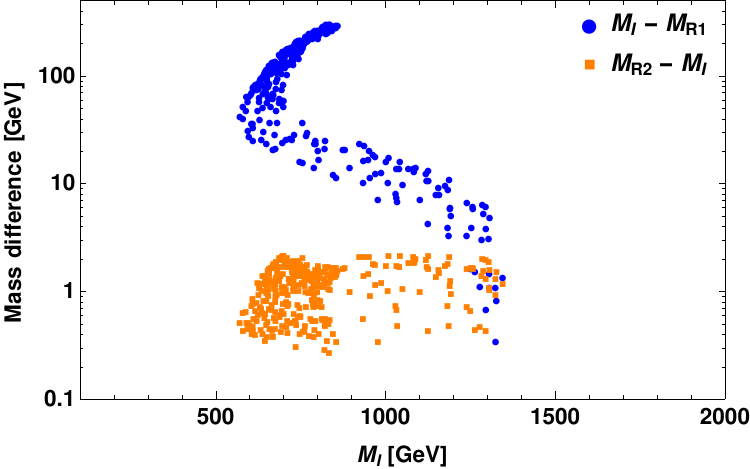}
 \caption{Top left panel displays relic density as a function of dark matter mass, illustrating the impact of $\delta$, with horizontal dashed lines corresponding to Planck bound \cite{Aghanim:2018eyx}.  Top right panel shows the parameter space consistent with aspects related to DM and neutrino in $M_{R1}-\delta$ plane. Lower panel projects the mass splitting of CP-even mass eigen states with CP-odd scalar.}
 \label{relic}
\end{figure}

Furthermore, we project relic density as a function of lightest dark matter mass ($M_{R1}$) in the upper left panel of Fig. \ref{relic} for the two benchmarks. Lower mass splitting basically induces new annihilation and co-annihilation channels to contribute, thereby giving larger cross section and smaller relic density. Hence, the curve (magenta) begins to meet Planck bound for larger DM mass, upper right panel portrays the parameter space in the plane of $M_{R1}-\delta$. Lower panel of Fig. \ref{relic} projects the mass splitting of CP-odd scalar with CP-even mass eigen states. As the splitting is clearly above $100$ KeV, large $Z$-portal WIMP-nucleon cross section is avoided \cite{Hambye:2009pw}. 
\\
%

\subsection{Oblique parameters and collider constraints}
\begin{figure}[htb!]
\centering
\includegraphics[scale=0.4]{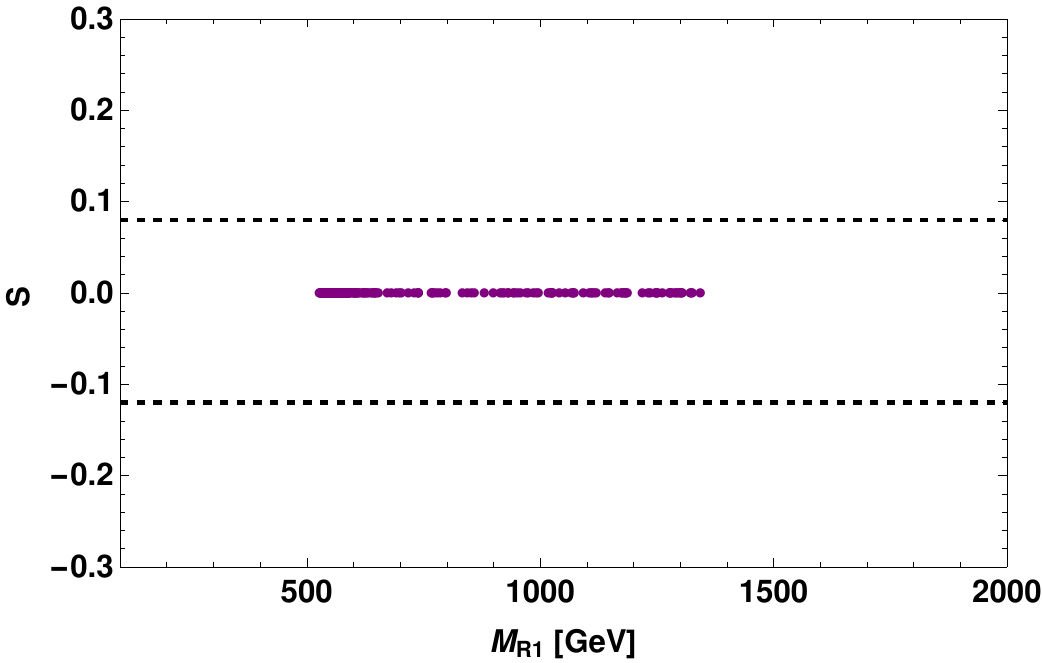}
\hspace{0.2 cm}
\includegraphics[scale=0.4]{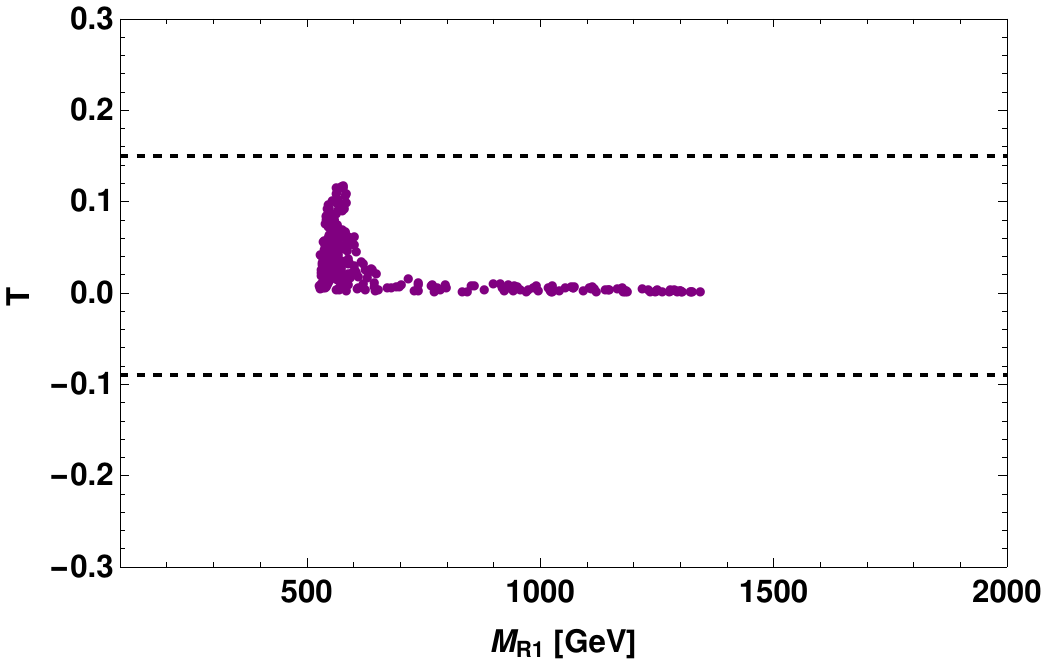}
\hspace{0.2 cm}
\includegraphics[scale=0.4]{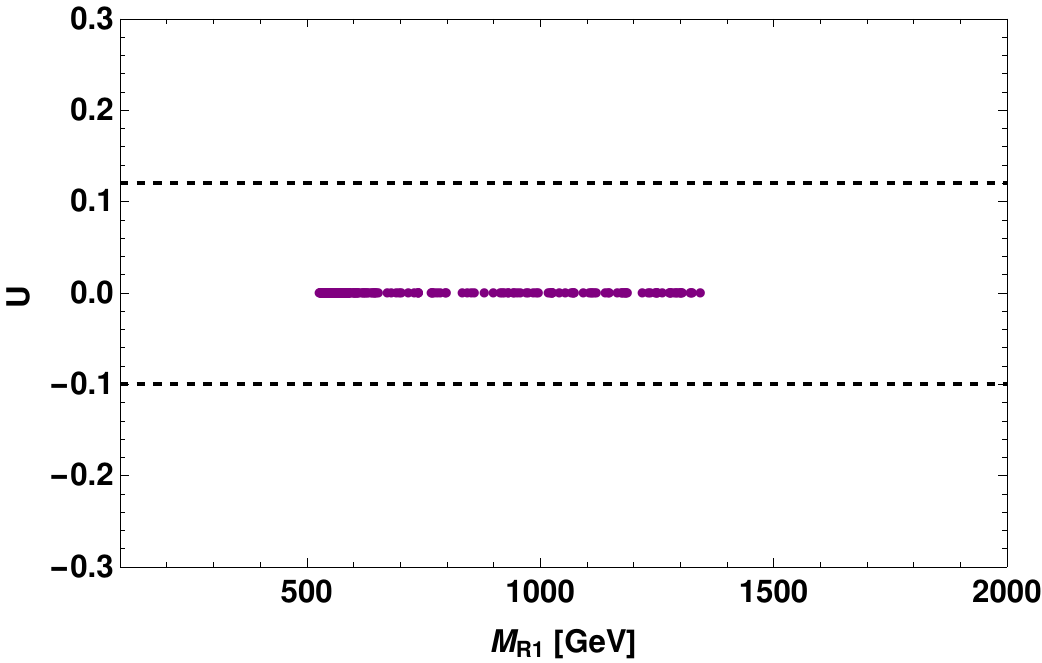}
 \caption{Oblique parameters computed for the parameter space consistent with dark matter and neutrino aspects. Horizontal dashed lines correspond to $1\sigma$ region of the current bounds from PDG \cite{ParticleDataGroup:2022pth}}
 \label{EWP}
\end{figure}

The masses of scalar multiplet components can get constrained from the self energies of gauge bosons and also collider studies as well. Precision tests can be made by looking at the oblique parameters i.e., $S,T$ and $U$. The contributions to these parameters from a scalar multiplet of hypercharge $Y$ is given by \cite{Lavoura:1993nq,Cheng:2022hbo}
\begin{eqnarray}
     S &=& -\frac{Y}{3\pi} \sum_{I_3=-I}^{+I} I_3 \ln\frac{M^2_{I_3}}{\mu^2} -\frac{2}{\pi}\sum_{I_3=-I}^{+I} (I_3 c_w^2 - Ys_w^2)^2 ~\xi\left(\frac{M_{I_3}^2}{M_Z^2},\frac{M_{I_3}^2}{M_Z^2}\right),\nn\\
     T &=& \frac{1}{16\pi c_w^2 s_w^2 M_Z^2} \sum_{I_3=-I}^{+I} (I^2-I_3^2+I+I_3) ~\Theta_+ (M_{I_3}^2,M^2_{I_3-1}),\nn\\
     U &=& \frac{1}{6\pi} \sum_{I_3=-I}^{+I} (I^2+I-2I_3^2) \ln\frac{M^2_{I_3}}{\mu^2} + \frac{1}{\pi} \sum_{I_3=-I}^{+I}\Bigg[2(I_3c_w^2 - Ys_w^2)^2 ~\xi\left(\frac{M_{I_3}^2}{M_Z^2},\frac{M_{I_3}^2}{M_Z^2}\right) \nn\\
    && -(I^2-I_3^2+I+I_3)~\xi\left(\frac{M_{I_3}^2}{M_W^2},\frac{M_{I_3-1}^2}{M_W^2}\right)\Bigg],
\end{eqnarray}
where, $I$ stands for weak isospin, $I_3$ corresponds to third component of isospin, $\mu^2$ stands for mass parameter for dimensional regularization and the functions used take the form \cite{Lavoura:1993nq,Cheng:2022hbo}
\begin{eqnarray}
	\xi(x,y)&=&\frac{1}{6}(x-y)^2 -\frac{5}{12} (x+y) + \frac{4}{9} +\frac{1}{4}\left[-\frac{1}{3}(x-y)^3 + x^2-y^2	-\frac{x^2+y^2} {x-y}  \right]\ln \frac{x}{y} \nonumber\\&-&\frac{1}{12}d(x,y)g(x,y)\;,\nonumber\\
	\Theta_+(x,y)&=&x+y-\frac{2xy}{x-y} \ln \frac{x}{y},
\end{eqnarray}
where $d(x,y)=-(x-y)^2+2(x+y)-1$,
\begin{eqnarray}
g(x,y)=\left\{
	\begin{array}{cl}
		-2\sqrt{d(x,y)}\left(\arctan \left[\frac{x-y+1}{\sqrt{d(x,y)}}\right]
		-\arctan \left[\frac{x-y-1}{\sqrt{d(x,y)}}\right]\right),\;\;\; &  \;\;\; d(x,y) > 0, \\
  0\;,\;\; &  \;\;\; d(x,y) = 0, \\
		\sqrt{-d(x,y)}\ln \left[\frac{x+y-1+\sqrt{-d(x,y)}}{x+y-1-\sqrt{-d(x,y)}}\right],\;\;\; &  \;\;\; d(x,y) < 0.
	\end{array} \right.
\end{eqnarray}
We compute these parameters for both the scalar triplets ($Y = 1,0$) and then write them in the mass basis of physical scalars using Eqn. \ref{flav to mass}. We calculate and project these parameters in Fig. \ref{EWP} for the parameter space consistent with dark matter and neutrino aspects. We notice that they are consistent with $1\sigma$ region of the current bounds on oblique parameters from PDG \cite{ParticleDataGroup:2022pth}. 

Moving on to collider limits, there exist constraints on the masses of triplet scalar components both neutral and charged from ATLAS \cite{ATLAS:2021jol}. The mass of doubly charged scalar ($\Delta^{++}$) is investigated through the production and associated production channels with $W^+W^-W^+W^-$ and $W^+W^-W^+Z$ in the final state respectively, where they exclude $230-350$ GeV mass range. However, these channels require triplet to obtain a non-zero vacuum expectation value and have decay channels for singly charged (to $WZ$ state) and doubly charged (to $WW$ state) scalars. Hence, these bounds are not applicable for the present model, since these decays are forbidden by $Z_2$ symmetry.
\section{$\nu$MM implications on XENONnT}
Recently XENONnT \cite{XENON:2022ltv} has released a new data with upgraded detector and a total exposure of $1.16$ ton-years and reduced systematic uncertainties. With new upgrade, more than 50$\%$ of background reduction has been achieved and unlike its predecessor XENON1T, no excess events were reported in $1-7$ keV energy range of electron recoil. In this paper, we use XENONnT data to derive model independent limits on transition and effective magnetic moments by examining the changes in the $\nu-e^-$ elastic scattering cross section at low energies. We use non-maximal $\theta_{23}$ mixing to distinguish between the muon and tau neutrino interactions. We consider XENONnT background without solar contribution and then add the expected events due to new physics such as neutrino magnetic moment.
In the presence of magnetic moment, the total differential cross section of $\nu-e^{-}$ scattering can be written as \cite{Giunti:2015gga}
\begin{eqnarray}
    \left( \frac{d \sigma}{d T_{r}} \right)_{\rm TOT} =  \left( \frac{d \sigma}{d T_{r}} \right)_{\rm SM} + \left( \frac{d \sigma}{d T_{r}} \right)_{\rm EM},
    \label{cross-section}
\end{eqnarray}
where, $T_{r}$ is the electron recoil energy. The first contribution in eq. \ref{cross-section} is due to standard weak interactions, given by
\begin{equation}
\left( \frac{d \sigma}{d T_{r}} \right)_{\rm SM} =  \frac{G_F^2 m_e}{2\pi} \left[(g_V + g_A)^2 + \left(1-\frac{T_{r}}{E_\nu}\right)^2(g_V - g_A)^2 + \left(\frac{m_e T_{r}}{E_\nu^2}\right)(g^2_A - g^2_V)\right].
\end{equation}
Here, $G_F$ stands for the Fermi constant, $E_{\nu}$ is the neutrino energy. $g_{V}$ and $g_{A}$ are the vector and axial vector couplings, which can be expressed in terms of weak mixing angle $\theta_W$ as 
\begin{eqnarray}
&&g_V = 2\sin^2 \theta_W + \frac{1}{2}, ~~~g_A = 1/2 ~~{\rm for}~~ \nu_e\;, \nn\\
&&g_V = 2\sin^2 \theta_W - \frac{1}{2},~~~ g_A = -1/2 ~~{\rm for}~~ \nu_\mu, \nu_\tau\;.
\end{eqnarray}
The second contribution in (\ref{cross-section}) comes from the effective electromagnetic vertex of the neutrinos, i.e.,  magnetic moment contribution, which can be expressed as
\begin{eqnarray}
    \left( \frac{d \sigma}{d T_{r}} \right)_{\rm EM} = \frac{\pi \alpha^{2}}{m^{2}_{e}} \left( \frac{1}{T_{r}} - \frac{1}{E_{\nu}} \right) \left( \frac{\mu_{\nu_{\alpha \beta}}}{\mu_{B}} \right)^{2}.
\end{eqnarray}
In the above, $\alpha$ is the fine-structure constant and $\mu_{\nu_{\alpha \beta}}$ is the neutrino magnetic moment and $\mu_{B}$ stands for Bohr Magneton. The differential event rate to estimate the XENONnT signal is given by
\begin{eqnarray}
\frac{dN}{ dT_{\rm vis}} &=&  n_{\rm e}\times \int^{E^{\rm max}_\nu}_{E^{\rm min}_\nu} dE_\nu \int_{T^{\rm min}_r}^{T^{\rm max}_r} dT_{r} \left(\frac{d\sigma^{\nu_e e}}{dT_{r}} \overline{P_{ee}} + \cos^2 \theta_{23} \frac{d\sigma^{\nu_\mu e}}{dT_{r}} \overline{P_{e\mu}} + \sin^2 \theta_{23} \frac{d\sigma^{\nu_\tau e}}{dT_{r}} \overline{P_{e\tau}}\right)  \nonumber\\
&& \times \frac{d\phi_s}{dE_\nu} \times \epsilon(T_{\rm vis})\times G(T_{r},T_{\rm vis})\;,
\label{nu-event}
\end{eqnarray}
where, $T_{\rm vis}$ represents the visible electron recoil energy at the detector, $\epsilon (T_{\rm vis})$ is the detector efficiency \cite{XENON:2022ltv} and $d\phi_s/dE_\nu$ is the solar neutrino flux \cite{Bahcall:2004mz}. $G(T_{r},T_{\rm vis})$ represents the normalized Gaussian smearing function, which takes into account the limited energy resolution of the detector with resolution power $\sigma (T_{\rm vis})/T_{\rm vis} = 0.0015 + (0.3171/\sqrt{T_{\rm vis} ~{\rm [keV]}})$. In the above, $\overline{P_{ee}}$ and $\overline{P_{e \mu / \tau}}$ are the length averaged neutrino disappearance and appearance oscillation probabilities in the presence of matter effect respectively, can be expressed as \cite{Khan:2022bel}
\begin{eqnarray}
    && \overline{P_{ee}} = \sin^4 \theta_{13} + \frac{1}{2} \cos^4 \theta_{13} \left( 1 + \cos 2\theta^{m}_{12} \cos 2\theta_{12} \right),\nn\\
    && \overline{P_{e \mu }} + \overline{P_{e \tau}}= 1- \overline{P_{ee}}\;,
\end{eqnarray}
where, $\theta^{m}_{12}$ is the effective mixing angle in the presence of matter effect \cite{Lopes:2013nfa} and we take the mixing angles from NuFit-5.2 \cite{Esteban:2020cvm}. 
In  eqn. (\ref{nu-event}), the differential cross sections can be expressed as the sum of SM and magnetic moment contributions as follows
\begin{eqnarray}
&&\frac{d\sigma^{\nu_e e}}{dT_{r}} = \left(\frac{d\sigma}{dT_{r}}\right)_{SM} + \left(\frac{d\sigma}{dT_{r}}\right)_{\mu_{\nu_{e \mu}}} + \left(\frac{d\sigma}{dT_{r}}\right)_{\mu_{\nu_{e \tau}}}, \nn\\
&&\frac{d\sigma^{\nu_{\mu} e}}{dT_{r}} = \left(\frac{d\sigma}{dT_{r}}\right)_{SM} + \left(\frac{d\sigma}{dT_{r}}\right)_{\mu_{\nu_{e \mu}}} + \left(\frac{d\sigma}{dT_{r}}\right)_{\mu_{\nu_{\mu \tau}}}, \nn\\
&&\frac{d\sigma^{\nu_{\tau} e}}{dT_{r}} = \left(\frac{d\sigma}{dT_{r}}\right)_{SM} + \left(\frac{d\sigma}{dT_{r}}\right)_{\mu_{\nu_{e \tau}}} + \left(\frac{d\sigma}{dT_{r}}\right)_{\mu_{\nu_{\mu \tau}}}.
\end{eqnarray}
The integration limits on $E_{\nu}$ goes from $\left(T_{r} + \sqrt{2 m_{e} T_{r} + T^{2}_{r}}\right)/2$ to 420 keV (corresponding to the upper limit of pp-chain in Sun).  The other limit $T_{r}$ describes the threshold of the detector, which runs from 1 keV to 140 keV (recoil energy of interest). We now estimate the neutrino transition magnetic moment using XENONnT data through the least-squared statistical method and define the following $\chi^2$ function, 
\begin{eqnarray}
    \chi^2 &=& \sum_{k=1}^{70} \left( \left[ \left( \frac{dN}{dT_{\rm vis}}(1+\alpha) + B\right)^{\rm theory}_{k} - \left(\frac{dN}{dT_{\rm vis}} \right)^{\rm observed}_{k} \right] \Bigg{/} \sigma_k \right)^2 
     + \left( \frac{\alpha}{\sigma_{\alpha}} \right)^2.
\end{eqnarray}
Here, the subscript $k$ represents the $k^{\rm th}$ bin of our theoretical prediction and observed events, $\sigma_{k}$ corresponds to the statistical uncertainty in each bin. We have considered the systematic error ($\sigma_{\alpha}$) to be around $10\%$ (reflected through the pull parameter $\alpha$), corresponding to the solar neutrino flux for our analysis. We also included the penalties for the uncertainties in mixing angles $\theta_{12}$, $\theta_{13}$ and $\theta_{23}$.
\begin{figure}
    \centering
    \includegraphics[scale=0.5]{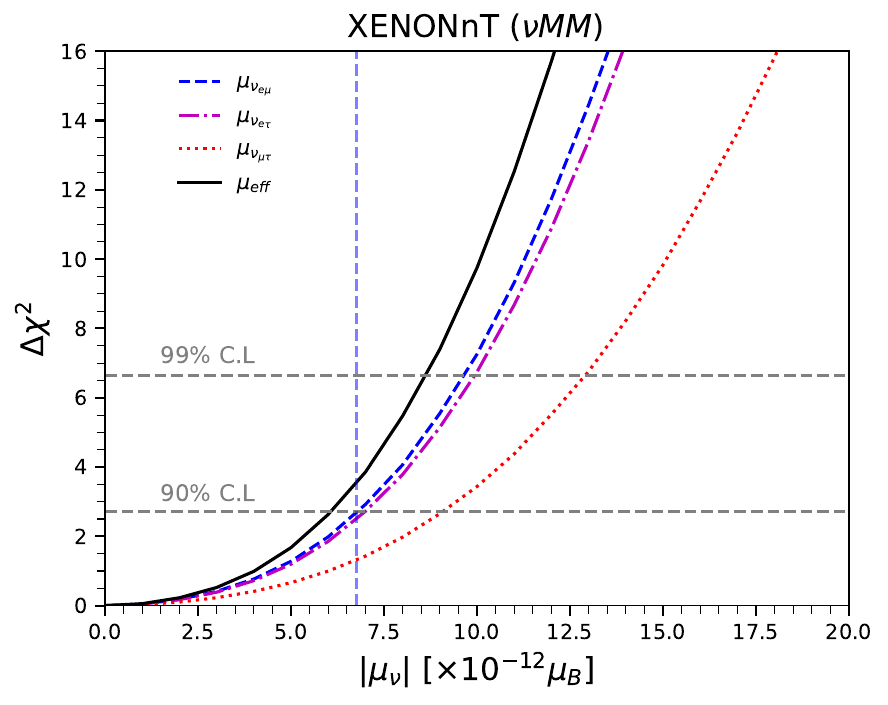}
    \includegraphics[scale=0.5]{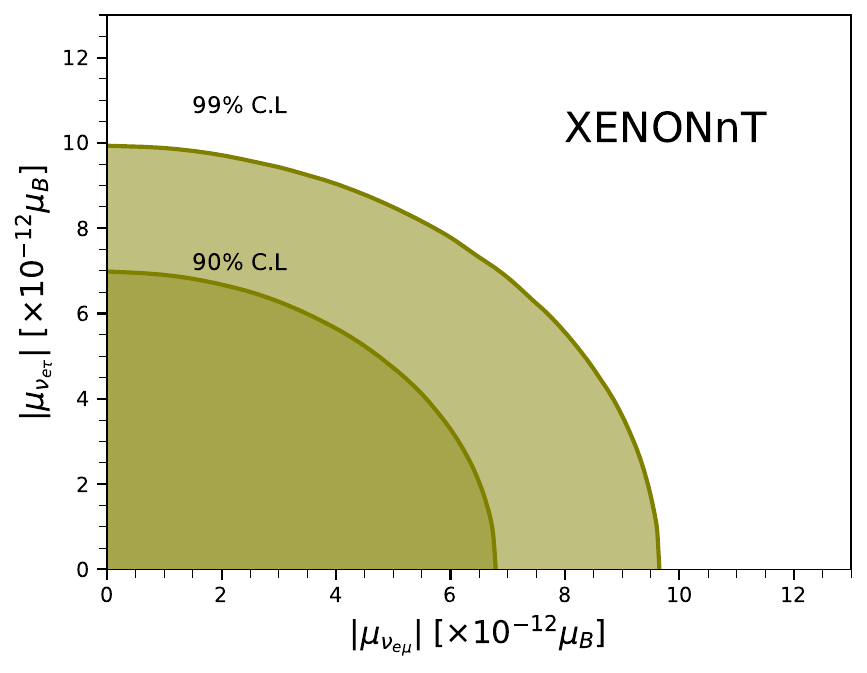}
    \caption{Left panel projects bounds on transition and effective magnetic moment, right panel shows the allowed region of transition magnetic moments in 90$\%$ and 99$\%$ C.L. for one degree of freedom at XENONnT experiment.}
    \label{bound}
\end{figure}
Left panel of Fig \ref{bound} shows the bounds on the neutrino transition magnetic moment and effective magnetic moment at 90$\%$ and 99$\%$ C.L. for the experiment XENONnT. The blue, violet and red curves represent the transition magnetic moment sensitivity of the components $\mu_{\nu_{e \mu}}$, $\mu_{\nu_{e \tau}}$ and $\mu_{\nu_{\mu \tau}}$ respectively, the black curve corresponds to the effective magnetic moment sensitivity of XENONnT experiment. The two grey horizontal lines stand for the sensitivity at 90$\%$ and 99$\%$ C.L. and the blue vertical line indicates the sensitivity of the experiment to transition magnetic moment $\mu_{\nu_{e \mu}}$ at 90$\%$ C.L. All the bounds on transition magnetic moments and effective magnetic moment are listed in Table \ref{table_bound}. As the Sun is the source of electron type of neutrinos, we notice that the bounds on transition magnetic moments $\mu_{\nu_{e \mu}}$ and $\mu_{\nu_{e \tau}}$ are more constrained than $\mu_{\nu_{\mu \tau}}$. The right panel of Fig \ref{bound} shows the allowed region of transition magnetic moments in $\mu_{\nu_{e \mu}} - \mu_{\nu_{e \tau}}$ plane at 90$\%$ and 99$\%$ C.L for the experiment XENONnT. Thus, we end the discussion by saying that our model successfully generates transition magnetic moments in the allowed region.

\begin{table}
\centering
\begin{tabular}{ |p{3cm}|p{3cm}|p{3cm}|  }
\hline
\multicolumn{3}{|c|}{$|\mu_{\nu}|$ $ [\times 10^{-12} \mu_{B}]$}  \\
\hline
XENONnT  &  90$\%$ C.L & 99$\%$ C.L\\ \hline
        $\mu_{\rm eff}$ & $<$ 6.08 & $<$ 8.6 \\ \hline
        $\mu_{\nu_{e \mu}}$ & $<$ 6.77 &  $<$ 9.63\\ \hline
        $\mu_{\nu_{e \tau}}$ &$<$ 6.98 & $<$ 9.94 \\ \hline
        $\mu_{\nu_{\mu \tau}}$ & $<$ 9.04 & $<$ 12.9\\ \hline        
\end{tabular}
\caption{Bound on neutrino magnetic moment at 90$\%$ and 99$\%$ C.L at XENONnT experiment.}
    \label{table_bound}
\end{table}

\section{Concluding remarks}
The primary motive of this model is to provide a simplified framework to invoke neutrino  magnetic moment along with dark matter. The trick is to realize neutrino electromagnetic vertex with dark matter running in the loop. Using vector-like fermion and scalar multiplets, we attain magnetic moment and mass for light neutrinos via Type-II radiative scenario. The scalar triplet components annihilate and co-annihilate through the Standard Model scalar and vector bosons to provide correct order of dark matter relic density in the Universe (consistent with Planck satellite) and also get recoiled from detector, giving spin-independent cross section which is sensitive to stringent upper limit (LZ-ZEPLIN). Both the neutrino and dark matter aspects are thoroughly discussed in a common model parameter space, illustrated with suitable plots and benchmark values. Using XENONnT data, we have put bounds on transition magnetic moments in a model independent way. Finally, we sign off by saying that the model stands simple but phenomenologically rich in providing a common platform to address neutrino properties (magnetic moment and mass) and also dark matter physics (relic density and direct detection).
\acknowledgments 
SS and RM would like to acknowledge University of Hyderabad IoE project grant no. RC1-20-012. DKS acknowledges the support of Prime Minister's Research Fellowship, Government of India.  DKS would like to convey thanks to Ms. Priya Mishra and Ms. Papia Panda for useful input.


\bibliography{nmm_DM}
\end{document}